\documentclass[12pt,preprint]{aastex}

\def\gapp{\ \lower 3pt\hbox{${\buildrel > \over \sim}$}\ }
\def\lapp{\ \lower 3pt\hbox{${\buildrel < \over \sim}$}\ }

\def\gapprox{\;\rlap{\lower 3.0pt                       
             \hbox{$\sim$}}\raise 2.5pt\hbox{$>$}\;}
\def\lapprox{\;\rlap{\lower 3.0pt                       
             \hbox{$\sim$}}\raise 2.5pt\hbox{$<$}\;}

\newcommand{\rn}[1]{(\ref{#1})}

\newcommand{\ben}{\begin{enumerate}}
\newcommand{\een}{\end{enumerate}}

\newcommand{\bc}{\begin{center}}
\newcommand{\ec}{\end{center}}
\newcommand{\bi}{\begin{itemize}}
\newcommand{\ei}{\end{itemize}}
\newcommand{\bt}{\begin{tabbing}}
\newcommand{\et}{\end{tabbing}}
\newcommand{\bd}{\begin{description}}
\newcommand{\ed}{\end{description}}

\newcommand{\be}{\begin{equation}}
\newcommand{\ee}{\end{equation}}
\newcommand{\bea}{\begin{eqnarray}}
\newcommand{\eea}{\end{eqnarray}}
\newcommand{\bean}{\begin{eqnarray*}}
\newcommand{\eean}{\end{eqnarray*}}

\newcommand{\ff}[2]{{\textstyle \frac{#1}{#2}}}

\shorttitle{Short-Period Terrestrial Planets}
\shortauthors{Mardling \& Lin}

\begin{document}

\title{On the Survival of Short-Period Terrestrial Planets}

\author{Rosemary A.\ Mardling}
\affil{Centre for Stellar and Planetary Astrophysics, School of Mathematical Sciences, Monash University, Victoria 3800, Australia}
\email{rosemary.mardling@sci.monash.edu.au}

\and

\author{D.\ N.\ C.\ Lin}
\affil{UCO/Lick Observatory, University of California, Santa Cruz, CA95064, USA}

\begin{abstract}
The currently feasible method of detection of Earth-mass planets is
transit photometry, with detection probability decreasing with a
planet's distance from the star. The existence or otherwise of
short-period terrestrial planets will tell us much about the planet
formation process, and such planets are likely to be detected first if
they exist. Tidal forces are intense for short-period planets, and
result in decay of the orbit on a timescale which depends on
properties of the star as long as the orbit is circular. However, if an eccentric
companion planet exists, orbital eccentricity ($e_i$, $i$ for inner orbit) is induced and
the decay timescale depends on properties of the short-period planet,
reducing by a factor of order $10^5 e_i^2$ if it is terrestrial.  Here
we examine the influence companion planets have on the tidal and
dynamical evolution of short-period planets with terrestrial
structure, and show that the relativistic potential of the star is
fundamental to their survival.

\end{abstract}


\section{Introduction}

In the past few years, the radial velocity method has been used to
detect more than 100 extrasolar planets \citep{marcy} with minimum
masses in the range 0.11 to 17 Jupiter masses. With the current lower
detection limit of 3 m/s, discovering Earth-mass planets with periods
longer than a few days, this way is untenable.  In principle,
Earth-mass planets with periods less than 3 days may be marginally
detectable with 1 m/s velocity precision \citep{narayan}.
But a large number of observations is needed to reduce the noise
associated with the stellar radial velocity ``jitter'' \citep{saar}.

Hopes for detecting Earth-mass planets in the near future lie with
transit photometry and microlensing. So far, four 
gas giants have been observed crossing
the faces of their stars \citep{brown,konacki03,konacki04,bouchy}, 
three of which were discovered using microlensing
techniques \citep{konacki03,konacki04,bouchy}.
While the transit of a gas giant results in a
maximum reduction of the star's brightness of around 1-2\%,
an Earth-mass planet would dim the star's light by only a few
parts in 10,000.
Nonetheless, such high-precision photometry is
currently feasible in the MOST satellite and will be employed in the
Kepler \citep{koch} mission.  Simple geometrical arguments give the
probability of seeing a planet cross the face of its parent star along
our line of sight as $R_*/2a$, where $R_*$ is the radius of the star
and $a$ is the distance of the planet from the star during the
transit. Thus it is most likely that short-period Earth-like planets
will be discovered first if they exist, especially since a weak signal
can be strengthened by the integration of many orbits.
This is not only true for planets discovered directly with transit photometry,
but also those discovered using microlensing techniques
since followup transit photometry is necessary for their confirmation and study.

In Section 2 we review the process of tidal damping, in Section 3 we
discuss the dynamical effect of the presence of a companion, and in
Section 4 we present a discussion.

\section{Orbit Evolution due to Tidal Damping}

During the formation epoch, tidal interactions between Jupiter-mass
protoplanets and their nascent disks can cause them to spiral in
toward their host stars \citep{goldtremaine,linpap}, a migration
scenario \citep{lin96} adopted to account for the origin of extrasolar
planets with periods of a few days \citep{mayor}. Other scenarios have
been suggested \citep{rasio,nag,murray}, all of which involve
significant excitation of orbital eccentricity. Once a planet is in
the vicinity of the star, however, tidal forces will tend to
circularize the orbit as well as synchronize the planetary spin
\citep{rasio96}. Earth-mass terrestrial planets may also be brought
close to their host stars via one or more of these processes and their
discovery (or otherwise) will shed much light on the planet formation
process.  For example, along their migration paths, gas giants induce
residual terrestrial planets to undergo orbital evolution through
their resonant and secular interaction. The detection of both
short-period gas giants and terrestrial planets around any host stars
would provide overwhelming support for the core-accretion scenario.

Short-period planets are also capable of raising a substantial tide on
their host stars.  In order to avoid noise introduced by stellar
jitters, the present planet-search campaigns focus on target stars
with quiet chromosphere and slow spin rates.  With one exception
\citep{butler}, all stars observed hosting short-period (``hot'')
Jupiters have spin periods considerably longer than the orbital period
of the planet.  Mature solar-type stars tend to be sub-synchronous
because stellar winds carry away spin angular momentum through
magnetic braking \citep{soderblom}.  Within these slowly spinning
stars, the induced tidal oscillations act to dissipate their acquired
tidal energy, the tidal response tends to lag the line of centers of
the star and planet, and the resulting torque tends to spin up the
star at the expense of orbital energy and angular momentum.

The simplest quantitative description for tides is the equilibrium
tide model \citep{goldsoter,hut,egg} which is adequate for analyzing
the tidal response of terrestrial planets. In gaseous planets and
stars, tidal perturbation can excite resonant responses which strongly
enhance the dissipation rate \citep{zahn}.  In radiative stars, the
response is mostly in the form of g-mode oscillations whereas in
fully convective envelopes of rapidly spinning gaseous planets and
stars, it is in the form of inertial waves \citep{ogilvie}.
Since we are primarily interested in the fate of the terrestrial 
planets in this paper, we adopt here the equilibrium tide model.

To leading order in the orbital eccentricity, the timescale,
$\tau_a$, for orbital decay is given by \citep{mardling} \be
\frac{1}{\tau_{a}}=\left(\frac{\dot{a}_i}{a_i}\right)_{\rm star}
+\left(\frac{\dot{a}_i}{a_i}\right)_{\rm planet},
\label{adot}
\ee where \be \left(\frac{\dot{a}_i}{a_i}\right)_{\rm star}
=-\frac{9}{2}\left(\frac{n_i}{Q_*'}\right)
\left(\frac{M_i}{M_*}\right) \left(\frac{R_*}{a_i}\right)^5
\left[1-\left(\frac{P_{\rm orb}}{P_{\rm spin}}\right)\right] \ee and
\be \left(\frac{\dot{a}_i}{a_i}\right)_{\rm planet}
=-\frac{171}{4}\left(\frac{n_i}{Q_i'}\right)
\left(\frac{M_*}{M_i}\right) \left(\frac{R_i}{a_i}\right)^5 e_i^2 \ee
are the contributions to orbital decay (or expansion) from tidal
dissipation within the star and planet respectively, and the
assumption has been made that the planet spins synchronously with the
orbital motion. Here $P_{\rm orb}$ and $P_{\rm spin}$ are the orbital
and stellar spin periods respectively, $a_i$ is the semi-major axis of
the orbit, $n_i$ is the mean motion (orbital frequency), $M_i$, $R_i$,
and $Q'_i$ are respectively the mass, radius and {\it modified}
Q-value of the planet, the latter containing information about damping
efficiency and rigidity \citep{goldsoter} (see Table 1), and $M_*$,
$R_*$, and $Q'_*$ are the corresponding values for the star. 
If the orbit is perfectly circular the planet makes no contribution to
orbital shrinkage and the orbital decay timescale depends entirely on
properties of the star. Otherwise the planet dominates the tidal decay
process, and this is particularly true if it is Earth-like.  In the
case of slow stellar rotation and small eccentricity, the ratio of
contributions to orbital decay from the planet and the star is
$12.5e_i^2$ for a Sun-Jupiter pair, $1.3\times 10^5e_i^2$ for a
Sun-Earth pair, and $2.2\times 10^5e_i^2$ for an M dwarf-Earth pair,
where we have used $Q'$-values from Table 1. The discrepancy between
the Jupiter system and the Earth systems is mostly due to the factor
$10^4$ difference in $Q'$-values of gas giants and terrestrials. Thus
even a small orbital eccentricity allows a planet to dominate the
tidal decay process if it has terrestrial structure, and can reduce
the decay timescale considerably.

The timescale, $\tau_e$, for eccentricity damping (or excitation) is
given by an expression similar to Eqn. 1 \citep{goldsoter,mardling}
except that it does not depend on $e_i$ to leading order in
$e_i$. Nonetheless, unless the star spins rapidly, the planet
dominates the tidal circularization process \citep{dobbs}. Table 2
lists orbital decay and eccentricity damping timescales, together with
other system parameters including transit probabilities for various
real and hypothetical single planet systems.  Included are the three
of the shortest period planets discovered to date as well as
HD209458. The values of $a_i$ for the three hypothetical systems
(Sun-Jupiter, Sun-Earth and M dwarf-Earth) were chosen to indicate how
close to the host star we can ever expect to find planets in these
types of systems. Note especially the steep dependence of $\tau_a$ on
the eccentricity when the planet is Earth-like.

\section{Eccentricity Excitation due to a Companion}
HD209458 is included in Table~2 as one of only two transit systems
discovered so far \citep{brown}, the data for which allows an estimate
of the planet radius of $1.35R_J$ which we use here to estimate
timescales. The planet's relatively large size can be accounted for by
inflation due to tidal dissipation of energy \citep{boden01} if
$e_i\sim0.03$ and $Q'_i=10^6$, with the corresponding orbital decay
timescale being comparable to the age of the Universe.  The
observationally estimated eccentricity is small ($<0.05$) but not zero
(Fischer, private communication), even though $\tau_e$ is smaller than
the estimated age of HD209458 (see Table 2). In order to resolve this
paradox, the existence of a second planet has been conjectured
\citep{boden01} which secularly excites the eccentricity \citep{MD}.

This eccentricity-excitation mechanism puts severe constraints on the
existence of systems containing short-period terrestrial planets with
companion planets. Figure 1a shows the evolution of the eccentricity
of an Earth-mass planet in a 2.3 day orbit ($a_i$=0.02 AU) around an M
dwarf of mass $0.2M_\odot$ which spins with a period of 40 days.  It
has a Jupiter-mass companion at a distance 0.7 AU which itself has an
eccentricity of 0.2. The initial eccentricity of the inner planet is
zero, however, it is excited by the presence of the companion to a
maximum value which depends on the strength of additional perturbing
accelerations which include the spin and tidal bulges of the star and
planet, as well as the relativistic potential of the star
\citep{mardling}.

In the absence of any of these accelerations, the inner eccentricity
is excited to a value of 0.015 (the black curve). Taking half this
value to represent the equilibrium inner eccentricity (see following
paragraphs), the orbital decay timescale would be 7.1 Gyr. However, if
all the perturbing accelerations listed above are included, the
maximum inner eccentricity is only 0.0035, corresponding to an orbital
decay timescale of 84.3 Gyr. This suggests that relativity plays a
major role in the survival of short-period terrestrial planets with
companions (given that Nature produces such systems in the first
place). In fact it is the contribution it makes to the apsidal advance
of the inner orbit which is responsible for the inhibition of the
growth of the inner eccentricity \citep{holman}.  The fractional
contribution the relativistic potential makes to apsidal advance
compared to that made by a companion planet is given by
$\gamma/(1+\gamma)$, where $\gamma=4Z^2(M_\ast/M_o)
(a_o/a_i)^3/(1-e_i^2)$, with $M_o$ the mass of the companion planet
and $Z=a_i n_i/c$ the ratio of the orbital speed of the inner orbit to
the speed of light \citep{mardling}.  For the case illustrated in
Figure 1a, the relativistic potential is responsible for 95\% of the
apsidal advance of the inner orbit, while in the case of Mercury's
orbit around the Sun, it contributes only 7\%.

Even more extreme is the case in which a short-period terrestrial
planet has an Earth-mass companion. For example, for a Sun-Earth-Earth
system with semi-major axes 0.03 AU and 0.5 AU, the relativistic
potential is responsible for 99.9\% of the apsidal advance of the
inner orbit. If the eccentricity of the outer planet is 0.3 the
orbital decay timescale is 702 Gyr, while in the absence of relativity
it would be only 0.47 Gyr.

Using equations which govern the secular evolution of the orbital
elements for a dissipationless point-mass coplanar system
\citep{mardling,MD,wu} and assuming small values for the
eccentricities and $a_i/a_o$ one can obtain an estimate for the
variation of the inner eccentricity, $\delta e_i$, which also holds
when the minimum eccentricity is zero at which point $e_i$ is
discontinuous as in Figure 1a. This estimate will vary slightly for
moderately non-coplanar systems, and does not apply to resonant
systems \citep{MD,novak}. If tidal dissipation is taken into account,
the system behaves as a damped autonomous system with the familiar
circulatory and libratory behaviour (Fig. 1b).  For a given set of
initial conditions the system evolves to a fixed point corresponding
to $\varpi_i-\varpi_o=2n\pi$, where $n$ is some integer and $\varpi_i$
and $\varpi_o$ are the longitudes of periastron of the inner and outer
orbits respectively, and a finite equilibrium inner eccentricity,
$e_i^{(\rm eq)}$, given approximately by \be e_i^{(\rm
eq)}=\frac{\ff{5}{8}(a_i/a_o)e_o}{|1-(M_i/M_o)\sqrt{a_i/a_o}+\gamma|}
=\delta e_i/2.
\label{emax}
\ee 
Note that as well as the relativistic potential, $\gamma$ may include
contributions from other perturbing accelerations such as tidal and
spin bulges.  Figure 1c shows the dependence of $e_i^{(\rm eq)}$ on
$M_i/M_o$ in the absence of relativistic and other perturbing
accelerations ($\gamma=0$).  The solid curves were obtained by
integrating the Newtonian equations of motion for point masses
(averaged over the inner orbit), while the dashed curves are given by
Eqn. 4.  Agreement is good except near the point corresponding to
$M_i/M_o=\sqrt{a_o/a_i}$.  The inclusion of relativistic effects
introduces an additional scale which is proportional to $m_i/a_i$, and
this is illustrated in Fig. 1d for the case $a_0/a_i=10$. Only in one
case is the discrepancy between the full solution and that given by
Eqn. 4 evident.

Figure 2 consists of grey-scale plots of the orbital decay timescale of
an Earth-mass planet orbiting a solar-mass star at 0.03 AU, as a
function of the eccentricity, $e_o$, and semi major axis, $a_o$, of a
Jupiter-mass perturber.  The equilibrium eccentricity is calculated
using the full governing equations \citep{mardling}, and this is then
used in Eqn. 1 to obtain the decay timescale.  The top panel is for
the case where no perturbing accelerations are included, while the
bottom panel includes the relativistic potential of the star. It is
clear that relativity allows many systems to survive which would
otherwise have suffered tidal destruction during the current lifetime
of the star.

\section{Discussion}

Of particular interest amongst short-period terrestrial systems will
be those with low-mass stars for which the habitable zone (planet
surface temperature in the range 0-100$^o$) is at a distance where the
transit probability is not negligible.  Table 2 lists several
hypothetical systems composed of a $0.2 M_\odot$ M dwarf and an
Earth-mass planet for which the habitable zone is around 0.04
AU. Planets in such close proximity to their host stars will be
tidally locked so that one side of the planet is never directly
heated. However, systems for which the habitable zone is further out
may have planets locked in other spin-orbit resonances. Mercury is
locked in a 3:2 spin-orbit resonance which relies on its permanent
slight departure from sphericity as well as its substantially
non-circular orbit \citep{goldpeale}.  Such planets would be heated
more evenly and hence be perhaps better candidates for detecting the
signatures of life in their atmospheres.

While the Q-values of putative short-period terrestrial-type planets
may turn out to be somewhat higher than the Earth's, either because of
the absence of oceans or because of the different temperature profiles
caused by the close proximity to the parent star, it is clear that a
star's general relativistic potential must play a major role in the
survival of such planets, and hence it is vital that it be included in
any studies of this problem.

\acknowledgements  We thank A. Cumming, K. Freeman, G. Novak, and P. Sackett for useful discussions. This work was supported by the Victorian Partnership for Advanced computing Expertise Program Grant Scheme, and by NASA through NAGS5-11779
under its Origins program, JPL 1228184 under its SIM program, and NSF
through AST-9987417.


\clearpage


\begin{figure}
\epsscale{0.35}
\plotone{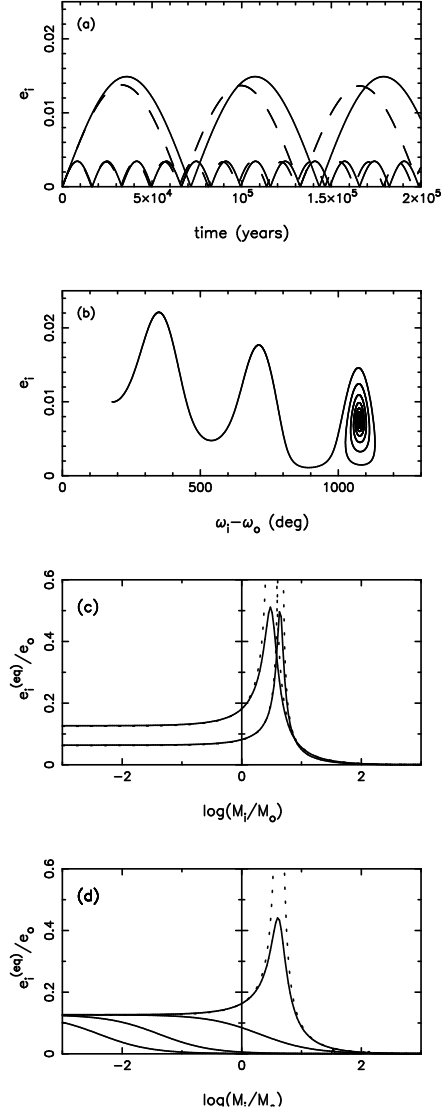}
\caption{\small Eccentricity excitation by a companion planet. 
{\bf a}, eccentricity evolution in the absence of tidal damping.
$M_*=0.2M_\odot$, $M_i=1 M_\oplus$, $M_o=1 M_J$, $a_i=0.02$ AU, $a_o=0.7$ AU, $e_o=0.2$,
and periastra initially aligned.
Solid curves: 
top=no perturbing accelerations; 
bottom=relativistic potential only.
Dashed curves:
top=spin and tidal bulge only;
bottom=both relativity and spin and tidal bulges.
{\bf b}, eccentricity evolution with tidal damping: circulatory and libratory behaviour in
the $e_i - (\varpi_i-\varpi_o)$ plane. System parameters
the same as in (a) for the case with no perturbing accelerations, but with initial $e_i=0.01$, 
periastra antialigned, and
an unrealistic $Q_i=0.2$
used to illustrate the behaviour. In general, a two-planet system will evolve to a constant $e_i$ with
periastra aligned as long as $\tau_e$ is shorter than the lifetime of the system.
{\bf c},  dependence of $e_i^{(\rm eq)}$ on $M_i/M_o$ with $\gamma=0$.
Solid curves: Newtonian point mass equations; dashed curves:  Eqn.~\rn{emax}.
Top set: $a_o/a_i=10$; bottom set: $a_o/a_i=20$.
{\bf d}, dependence of $e_i^{(\rm eq)}$ on $M_i/M_o$ with relativistic effects included.
$a_0/a_i=10$.
Bottom to top:
($a_i$/AU, $m_i/M_J$)=(0.05, 0.0033), (0.5, 0.0033), (0.05, 1), and (0.5,1).}
\label{figure1}
\end{figure}

\begin{figure}
\epsscale{0.6}
\plotone{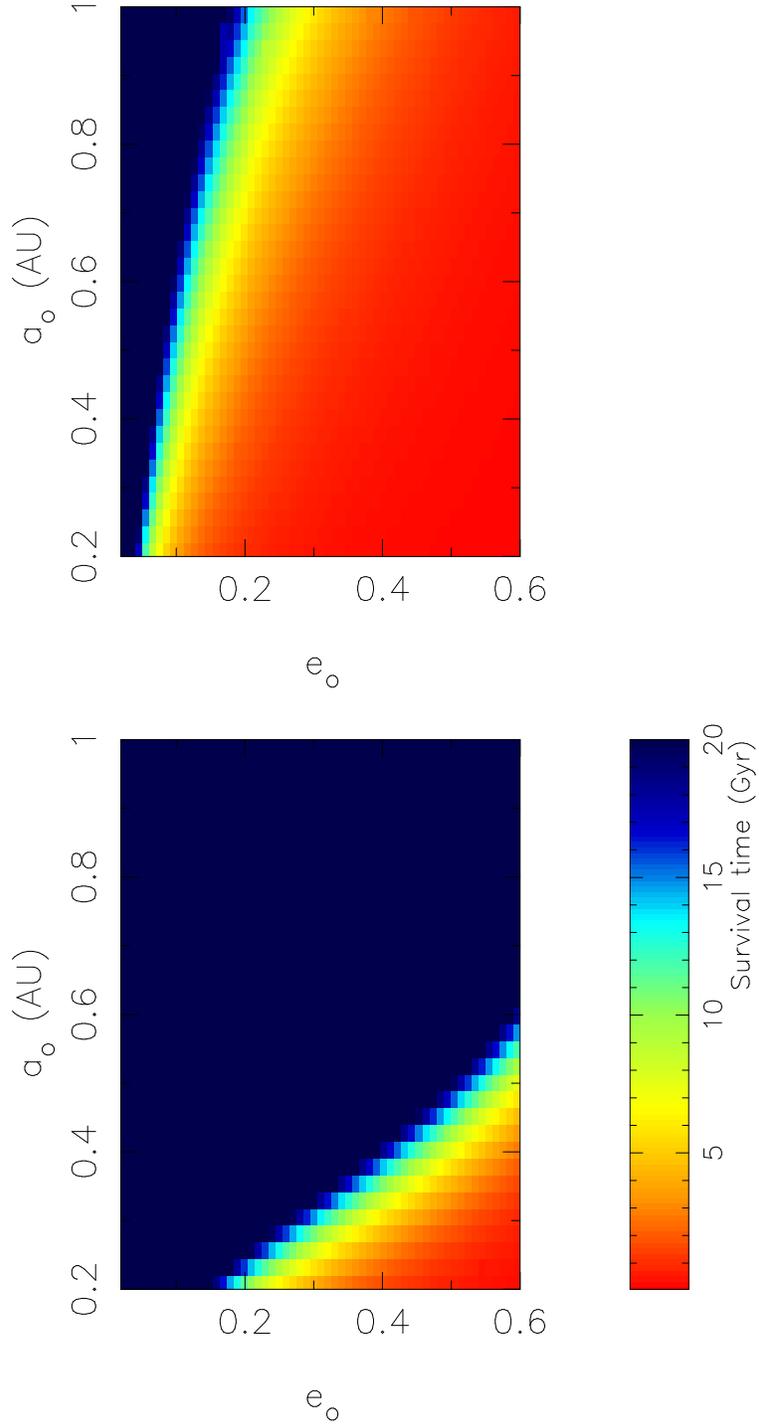}
\caption{Orbital decay timescales with and without relativity as functions
of the outer eccentricity, $e_o$, and semimajor axis, $a_o$.
$M_*=1 M_\odot$, $M_i=1 M_\oplus$, $M_o=1 M_J$, $a_i=0.03$ AU.
Top panel: no relativity. Bottom panel: with relativity.}
\label{figure2}
\end{figure}



\clearpage

Table 1: {\bf Structural parameters for various bodies}

\begin{tabular}{@{}llllc}
&&&&\\
Body & $k_L$ & $Q$ & $Q'$ & reference \\ \hline
&&&&\\
Venus & 0.295 & 40 & 203.4 & (1)\\
Earth & 0.3 & 21.5 & 107.5 & (2)\\
Mars & 0.14 & 86 & 921 & (3)\\
&&&&\\
Jupiter & 0.34 & $2.2\times 10^5$ & $10^6$ & (4)\\
&&&&\\
Sun & 0.02 & $1.3\times 10^4$ & $10^6$ & (5)\\
M dwarf & 0.30 & $2.0\times 10^5$ & $10^6$ & (5)\\
&&&&\\ \hline
\end{tabular}

\noindent Love numbers ($k_L$), specific dissipation functions ($Q$), and modified Q-values (Q'),
with $Q'=3Q/2k_L$ (Goldreich \& Soter, 1966, Murray \& Dermott, 1999).
Love numbers for gaseous bodies assume polytropic structure
with polytropic indices 1, 3 and 1.5 for Jupiter, Sun and 
M dwarf respectively. References. (1) Yoder, 1995a, (2) Dickey et al., 1994, (3) Yoder, 1995b,
(4) Yoder \& Peale, 1981, (5) Mathieu, 1994.

\newpage

\renewcommand{\baselinestretch}{1.}

\footnotesize

\hspace{-0.5cm}{Table 2: {\bf Orbital decay timescales}

\hspace{-0.5cm}\begin{tabular}{@{}lllllllllll}
&&&&&&&&&\\
System & $M_*/M_\odot$ & $M_i/M_J$ & $P_{\rm spin}$ &  $a_i$ (AU) &  $P_{\rm orb}$& Pr & $\tau_{e}$ & $\tau_a(0)$ & $\tau_a(0.01)$& $\tau_a(0.1)$ \\ \hline
&&&&&&&&&\\
Sun/Jupiter & 1.00 & 1.00 & 28 & 0.02 & 1.04 d &0.12& 0.03 &0.16 & 0.16 & 0.06 \\
                   &      &  &      & 0.03  &  1.91 d &0.08& 0.45 & 2.34  & 2.30 & 0.88 \\
                   &      &  &      & 0.04  & 2.93 d &0.03& 2.90 &15.8 & 15.6 & 5.83\\
&&&&&&&&&\\
Sun/Earth & 1.00 & 0.003 & 28 & 0.01 & 8.8 hr &0.23& $2.2-5$ & 0.53 & 0.009 & $8.1-5$ \\
                   &      &  &      & 0.015 & 16.2 hr &0.15& $3.0-4$ & 7.41 & 0.12 & $0.001$\\
                   &      &  &      & 0.02 & 25.0 hr  &0.12& $0.002$ &48.7 & 0.78 & 0.007\\
                   &      &  &      & 0.03 & 1.9 d  &0.08& $0.027$ &$\infty$ & 10.9 & 0.10\\
                   &      &  &      & 0.05 & 4.1 d  &0.05& $0.75$ &$\infty$ & $\infty$ & 2.82\\
&&&&&&&&&\\
M dwarf/Earth & 0.2 & 0.003 & 40 & 0.007 &  10.3 hr  &0.10& $2.4-5$ & 0.97 & 0.01 & $8.9-5$ \\
                   &      &  &      & 0.01 & 17.6 hr &0.07& $2.4-4$ & 9.88 & 0.097 & $9.0-4$\\
                   &      &  &      & 0.012  & 23.1 hr &0.06& $7.9-4$ &34.4 & 0.32 & 0.003\\ 
                   &      &  &      & 0.02  & 2.32 d &0.03& $0.02$ &$\infty$ & 8.82 & 0.082\\ 
                   &      &  &      & {\bf 0.04}  &  {\bf 6.56 d} & {\bf 0.01}&  {\bf 1.97} & {\bf $\infty$} &  
{\bf $\infty$} & {\bf 7.39} \\ 
&&&&&&&&&\\ 
OGLE-TR-56 & 1.04 & 0.9 & - & 0.023 & 1.212 d & 0.11 & 0.02 & 0.40 & 0.37 & 0.05 \\
HD83443 & 0.79 & 0.41 & 36.5 & 0.038 & 2.986 d &0.05 & 2.1 & 73.2 && \\
HD46375 & 1.00 & 0.25 & -       & 0.041 & 3.024 d &0.06& 5.1 &26.7 &&\\
%
&&&&&&&&&\\
{\bf HD209458}  & {\bf 1.05} & {\bf 0.66} & {\bf -} & {\bf 0.05} & {\bf 3.525 d} &{\bf 0.05}& {\bf 0.84} 
& {\bf 59.8} & {\bf 51.9} &{\bf 3.43}\\
&&&&&&&&&\\
\hline
\end{tabular}

\renewcommand{\baselinestretch}{1.2}

\normalsize

\noindent Orbital decay and eccentricity damping timescales (Gyr), and transit probabilities (Pr),
for various real and hypothetical single planet systems.
$M_i/M_J$ is the minimum mass of the planet in units of the mass of Jupiter, 
the actual mass being a factor $(\sin i)^{-1}$ higher when $i$ is the inclination of the orbit
to our line of sight,
`d' stands for days, and the argument of $\tau_a$ is
the eccentricity.
For all stars we took $Q'_*=10^6$,
for the known systems and the hypothetical Jupiter systems
$Q'_i=10^6$, while for the hypothetical Earth systems $Q'_i=21.5$. The symbol $\infty$
corresponds to $\tau_a>10^{11}$ yr. Entries in boldface for the last M dwarf-Earth system
correspond to the approximate {\it habitable zone} for such a star.

\end{document}